\documentclass[aip,jap,preprint]{revtex4-1}

\usepackage{graphicx}
\usepackage{amsmath}
\usepackage{amssymb}
\usepackage{natbib}
\usepackage{bm}
\usepackage{color}
\usepackage{siunitx}

\begin{document}
\title{Stabilization of vapor-rich bubble in ethanol/water mixtures and enhanced flow around the bubble}

\author{Mizuki Kato}
\affiliation{Department of Micro Engineering, Kyoto University,
Kyoto Daigaku-Katsura, Nishikyo-ku, Kyoto 615-8540, Japan.}

  \author{Kyoko Namura}
  \email{namura.kyoko.2s@kyoto-u.ac.jp}
\affiliation{Department of Micro Engineering, Kyoto University,
Kyoto Daigaku-Katsura, Nishikyo-ku, Kyoto 615-8540, Japan.}
  
  \author{Shinya Kawai}
\affiliation{Department of Micro Engineering, Kyoto University,
Kyoto Daigaku-Katsura, Nishikyo-ku, Kyoto 615-8540, Japan.}

 \author{Samir Kumar}
  \affiliation{Department of Electronics and Information Engineering,
 Korea University, Sejong 30019, Republic of Korea.}

 \author{Kaoru Nakajima}
 \author{Motofumi Suzuki}

\affiliation{Department of Micro Engineering, Kyoto University,
Kyoto Daigaku-Katsura, Nishikyo-ku, Kyoto 615-8540, Japan.}

\begin{abstract}
This study investigates the behavior of microbubbles generated by the local heating of an ethanol/water mixture and the surrounding flow. The mixture is photothermally heated by focusing a continuous-wave laser on a FeSi$_2$ thin film. Although the liquid is not degassed, vapor-rich bubbles are stably generated in an ethanol concentration range of 1.5--50 wt\%
 The vapor-rich bubbles absorb the air dissolved in the surrounding liquid and exhale it continuously as air-rich bubbles $\sim$ 1 \si{\um} in diameter. For the same ethanol concentration range, the solutal-Marangoni force becomes dominant relative to the thermal-Marangoni force, and the air-rich bubbles are pushed away from the high-temperature region in the fluid toward the low-temperature region. Further, it was experimentally demonstrated that Marangoni forces do not significantly affect the surface of vapor-rich bubbles generated in ethanol/water mixtures, and they produce a flow from the high-temperature to the low-temperature region on the vapor-rich bubbles, which moves the exhaled air-rich bubbles away from the vapor-rich bubbles near the heat source. These effects prevent the vapor-rich and exhaled air-rich bubbles from recombining, thereby resulting in the long-term stability of the former. Moreover, the flow produced by the vapor-rich bubbles in the non-degassed 0--20 wt\% ethanol/water mixture was stronger than that in degassed water.
The maximum flow speed is achieved for an ethanol concentration of 5 wt\%, which is 6--11 times higher than that when degassed water is utilized. The ethanol/water mixture produces vapor-rich bubbles without a degassing liquid and enhances the flow speed generated by the vapor-rich bubbles. This flow is expected to apply to driving and mixing microfluids.
\end{abstract}


\maketitle

\section{INTRODUCTION}

The demand for miniaturized heat exchangers has recently increased with the miniaturization of power devices. \cite{vanErp2020,Li2023} In fact, heat-dissipating fins utilized in heat exchangers are becoming smaller, with channel widths as small as 10--100 \si{\um}. A smaller channel width increases the surface area of the heat-dissipating fins. The cooling efficiency can be increased if the refrigerant can flow at high speed through these narrow channels. However, achieving high-velocity flow in a narrow channel is difficult because the fluid does not slip on the channel wall, and the flow velocity slows near the wall.
To overcome this problem, we focused on the flow generated by microbubbles. 
Bubbles exhibit large volume changes upon heating or ultrasound irradiation because of the gas phase's phase change and high compressibility. Such dynamics produce sound waves and ambient fluid flow, which are employed for fluid pumping \cite{Marmottant2003,Ahmed2016}, small-scale object manipulation \cite{Solovev2009,Xie2016,Khezri2020,Feng2023,Lee2023}, microsurgery \cite{Prentice2005}, ultrasound imaging \cite{Zhao2023} and ultrasound wavefront manipulation. \cite{Ma2020}
In addition, surface tension gradients can be induced at the bubble surface to generate shear forces known as Marangoni forces. Surface tension varies depending on the temperature and solute concentration. Therefore, a surface tension gradient was induced by applying these gradients to the bubbles. Marangoni forces have been reported as suitable for driving fluids \cite{Amador2019,Jones2020} and objects \cite{Dai2019,Lee2020,Wang2022} in microchannels. Because bubbles have a wide range of applications at the micrometer scale, there has been considerable interest in how they move under temperature and concentration gradients. \cite{Hu2022,LiXiaolai2023}
Moreover, previous studies have reported on the formation process and behavior of bubbles on local heating points at a micrometer scale.\cite{WangY2017,Wang2018,LiXiaolai2019,Zaytsev2020}
To date, the flows produced by bubbles on the micrometer scale have generally been reported to be 1--10 mm s$^{-1}$ on average, in time.
However, flow velocities in the order of 1--10 m s$^{-1}$ are required to achieve sufficient cooling effects in micro-water cooling. \cite{Missaggia1989}

Recently, we enhanced the flow around microbubbles to the order of 1 m s$^{-1}$. \cite{Namura2017,Namura2019,Namura2020,Namura2022} 
We adopted the photothermal conversion properties of the thin films to locally heat degassed water to produce water vapor-rich bubbles.
Thin films made of gold nanoparticles or FeSi$_2$ with thicknesses of 10--100 nm absorb light at wavelengths in the visible to near-infrared region. Most of the absorbed light was converted into heat. Therefore, by focusing a laser beam onto a thin film, the focal point can be utilized as a mobile point heat source. 
The local heating of non-degassed water produces bubbles composed mainly of air. \cite{Baffou2014} 
When a bubble is generated, the interior of the bubble is initially rich in water vapor, but the bubble quickly absorbs air molecules dissolved in the water.
Local heating of the surfaces of such air-rich bubbles induces Marangoni forces, which are surface tension gradients, and Marangoni convection on the order of 1--10 mm s$^{-1}$ can be generated. \cite{Namura2015,Namura2016,Namura2016JNP}
However, when degassed water is locally heated, a bubble of about 10 \si{\um} in diameter, mainly composed of water vapor, is generated.\cite{Namura2017} Interestingly, strong convection currents are generated around the bubble, reaching 1 m s$^{-1}$ just around the bubble. This phenomenon is useful for driving the fluid in the microchannel. 
It should be noted that laser irradiation is not necessarily required for this phenomenon to occur. Laser heating is used in our study because it is an easy method to control the heat generation density distribution for understanding the phenomenon. If a similar heat generation density distribution can be achieved on a cooling object, e.g., by patterning the thermal conductivity of the material, vapor rich bubbles and flows can be generated in the same way. 
However, to stably generate water vapor-rich bubbles, it is necessary to keep the water degassed. 
In particular, the narrower the channel, the greater the ratio of the surface area of the channel to the volume of water, so it is not easy to isolate water from non-condensable gases such as air.
This is a major drawback in the application of this phenomenon. 
In addition, the flow speed rapidly decreased with $-$3 to $-$2 power at a distance away from the bubble. Thus, a further enhancement of the flow speed is desired for cooling.

Here, we attempted to determine the conditions under which water vapor-rich bubbles could be generated stably without degassing. Li\cite{Li2017} and Nguyen\cite{Nguyen2019,Nguyen2019G} reported that water vapor-rich bubbles could be generated by the local heating of non-degassed water above a certain heat generation density. In non-degassed water, the bubble always takes in air molecules dissolved in the surrounding water.
If the air taken in is sufficiently small, the bubbles will oscillate over the local heating point, similar to vapor-rich bubbles in degassed water. \cite{Li2017,Namura2020}
Furthermore, if the heating density is sufficiently high and the bubbles oscillate violently, they can exhale the air they enter.\cite{Li2017} The bubble constantly takes in air molecules dissolved in the water, but each time the bubble collapses, the air taken in is exhaled as a small air-rich bubble, maintaining the water vapor rich condition in the bubble at the local heating point.
Although this phenomenon has potential applications in micro water cooling,
it is not optimal for cooling when the temperature of the object to be cooled rises slowly.
This is because air-rich bubbles form and cover the heat source at low heating densities, preventing the formation of vapor-rich bubbles that create a strong flow.
It would be beneficial if a similar phenomenon could be realized even at lower heating densities.
Recently, we reported an interesting phenomenon observed during the formation of air-rich bubbles at low heating densities in non-degassed water \cite{Namura2024}. 
Local heating of non-degassed water with a laser focused on a photothermal conversion thin film first produced a vapor-rich bubble with a radius of 9 \si{\um}. The bubbles oscillated and intermittently exhaled air-rich bubbles with radii of less than 1 \si{\um}. The exhaled air-rich bubbles initially move away from the vapor-rich bubbles; however, when the air-rich bubbles fused with each other and became larger, they were attracted to the vapor-rich bubbles. Eventually, these bubbles fused together, and the vapor-rich bubble on the heat source was simultaneously fed with a significant number of air molecules. Subsequently, air-rich bubbles began to grow on the heat source. The round-trip motion of the air-rich bubbles around the vapor-rich bubble on the heat source was explained by the balance between the thermal-Marangoni force and quasi-stationary drag force acting on the bubbles. As the bubbles enlarged, the Marangoni force acting on the bubbles became stronger, attracting the bubbles toward the heat source. This suggests that if the Marangoni force acting on the bubbles can be reduced or changed in direction, the round-trip motion of the air-rich bubbles can be suppressed, and the vapor-rich bubbles can be held stable.

To change the direction of the Marangoni forces, we focused on alcohol/water mixtures. In our previous study, we locally heated air-rich bubbles in water-alcohol mixtures and investigated the direction of the Marangoni convection around the bubbles \cite{Namura2018}. When the bubble was locally heated, not only a temperature gradient, but also a concentration gradient of alcohol due to the temperature gradient is generated on the bubble surface. Therefore, in the alcohol/water mixture, thermal-Marangoni forces due to the difference in surface tension caused by the temperature difference and solutal-Marangoni forces due to the difference in surface tension between water and alcohol act on the bubble surface. 
For example, in an ethanol/water mixture, thermal-Marangoni forces predominate at ethanol concentrations of 0--1.5 wt\% and 50--100 wt\%, and the total Marangoni force is directed from the high-temperature region to the low-temperature region. However, at 1.5--50 wt\%, solutal-Marangoni forces predominate, and the total Marangoni force is directed from the low-temperature region toward the high-temperature region. Therefore, we hypothesized that this phenomenon can be used to suppress the attraction of air-rich bubbles to a heat source.
Although studies on the formation of bubbles by heating alcohol/water mixtures and the behavior of these bubbles have been widely reported, most have used heat sources on the millimeter scale or larger \cite{Hovestreijdt1963,Abe2005,Abe2006,Zhou2014,Hu2019}.
In recent years, several studies have reported on the local heating of alcohol-water mixtures \cite{Butzhammer2017,Detert2020,Li2020}; however, systematic studies on convection around bubbles are still lacking.

In this study, we investigate the effect of the ethanol/water mixture concentration on the stabilization of vapor-rich bubbles and surrounding flow speed. 
The photothermal conversion properties of the FeSi$_2$ thin film were used to locally heat an ethanol/water mixture and generate bubbles.
We determined the concentration at which vapor-rich bubbles were stabilized by systematically changing the concentration of the ethanol/water mixture. 
In addition, we experimentally investigated the behavior of air-rich bubbles exhaled from vapor-rich bubbles and the flow produced by vapor-rich bubbles. Based on these results, the mechanism of vapor-rich bubble stabilization is qualitatively discussed in relation to the balance between the thermal-Marangoni, solutal-Marangoni, and quasi-steady drag forces. Furthermore, we report that vapor-rich bubbles generated in an ethanol/water mixture produce faster flows than vapor-rich bubbles generated in degassed water.

\section{Results and discussion}
\begin{figure*}[tbp]
\centerline{\includegraphics[bb = 0 0 563 362, width=12cm]{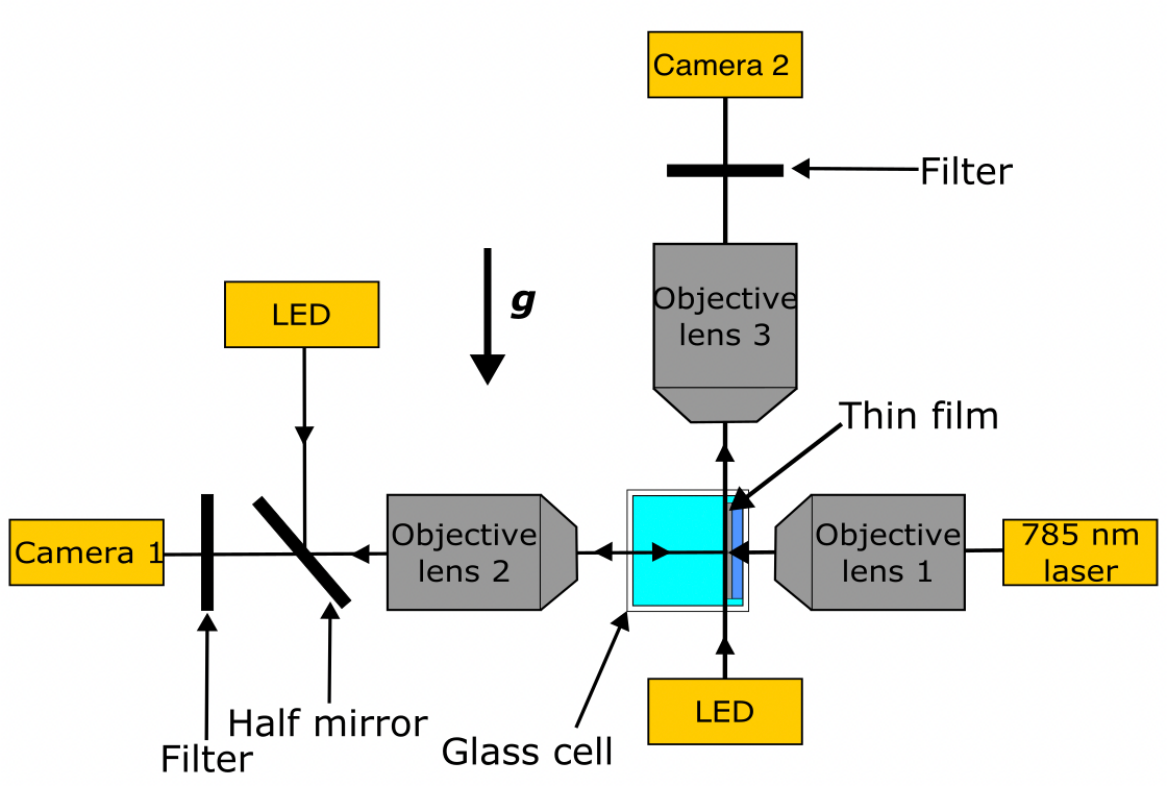}}
\caption{Schematics of experimental setup. We focused 785-nm-wavelength laser on FeSi$_2$ thin film to generate bubble. Shape and size of laser spot on thin film were adjusted with Camera 1. Bubble and fluid motion was observed by Camera 2.}
 \label{fig1}
\end{figure*}

\begin{figure*}[tbp]
\centerline{\includegraphics[bb = 0 0 761 630, width=15cm]{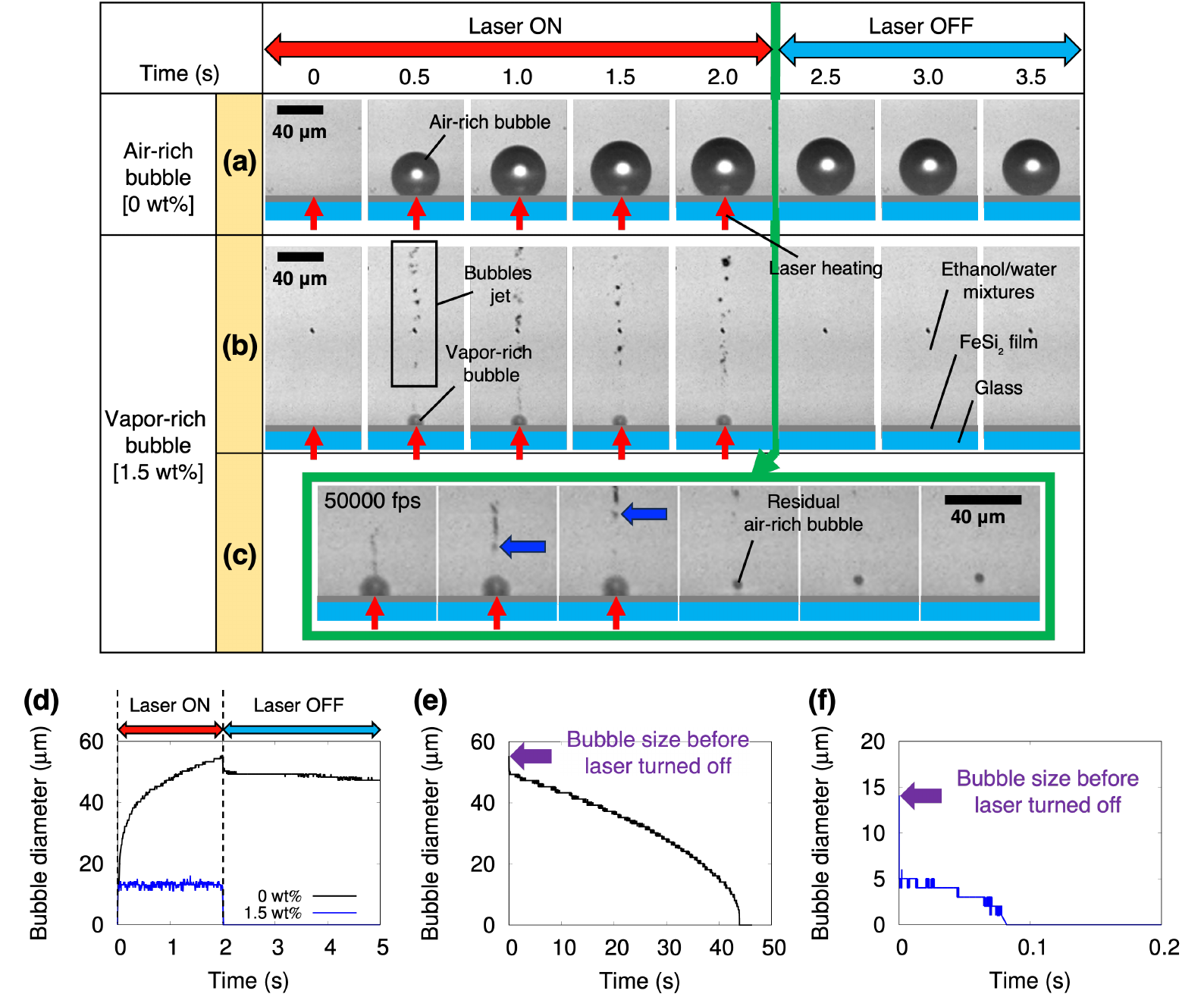}}
\caption{Typical time series microscope images of bubbles observed in (a) 0 wt\% and (b) 1.5 wt\% ethanol/water mixtures, irradiated with a laser for 2 seconds. (c) Bubble images before and after the cessation of laser irradiation in 1.5 wt\% ethanol/water mixture; images were taken at 50,000 fps. (d) Black and blue lines are the time variation of bubble diameter shown in (a) and (b), respectively.
(e) and (f) are the extinction processes of the bubbles in (a) and (b), respectively, after the laser irradiation is stopped.}
 \label{fig2}
\end{figure*}

The microbubbles generated by the photothermal conversion effect of a FeSi$_2$ thin film were observed with the experimental setup illustrated in Figure \ref{fig1} (see Experimental Section for details). 
Briefly, the 785-nm-wavelength laser was focused on the FeSi$_2$ thin film through objective Lens 1 to generate bubbles. The shapes and sizes of the laser spots on the thin film were adjusted with Camera 1. The generated microbubbles were then observed by Camera 2 from a direction parallel to the FeSi$_2$ thin-film surface. 
Figure \ref{fig2}(a--c) presents typical microscopic images of the bubbles observed in ethanol/water mixtures.
The bubbles were generated by irradiating a 27-mW laser beam for 2 s on an FeSi$_2$ thin film in ethanol/water mixtures.
Figure \ref{fig2}(a) illustrates a typical bubble containing mainly air generated in a 0 wt\% ethanol/water mixture, namely, non-degassed pure water. Here, such a bubble is called an "air-rich bubble." Photothermal conversion occurred at the laser spot on the FeSi$_2$ film, and microbubbles were generated on the spot. 
The black circles, with diameters of 45--55 \si{\um}, represent the air-rich bubble. The black line in Figure \ref{fig2}(d) shows the time evolution of the bubble diameter in Figure \ref{fig2}(a).
When the air-rich bubble was generated, its diameter increased to 55 \si{\um} within 2 seconds of laser irradiation. When the laser irradiation was stopped, the bubble diameter shrank by 7 \% within 0.01 s. 
After instantaneous shrinkage due to cooling caused by heating cessation, the bubble diameter decreases relatively slowly with time.
Figure \ref{fig2}(e) shows the time variation of bubble diameter until the bubble disappears after the laser irradiation is stopped.
The bubble contracted slowly over a period of about 45 seconds. This relatively long bubble lifetime occurs because air diffusion dominates the bubble shrinkage, which is the best evidence that the inside of the bubble is air-rich. \cite{Baffou2014,Zaytsev2020} 
Considering the change in the diameter of the bubble before and after the laser irradiation is stopped, taking into account the saturated vapor pressure and Laplace pressure, it can be estimated that about 80\% of the molecules inside the bubble during laser irradiation are air.
Here, we assumed the ambient pressure to be 1 atm and water temperature to be 25 \si{\degreeCelsius} when the laser irradiation is stopped. The detailed calculation methods can be found in Ref. [\cite{Hiroshige2024}].

In contrast, Figure \ref{fig2}(b) illustrates a typical bubble containing mainly water and ethanol vapor. Here, such a bubble is called a "vapor-rich bubble." Here, vapor-rich bubbles were frequently observed in the ethanol/water mixtures at specific concentrations, even when the liquid was not degassed. 
In Figure \ref{fig2}(b), the black circle with a diameter of approximately 13 \si{\um} on the laser spot represents a vapor-rich bubble.
The blue line in Figure \ref{fig2}(d) shows the time evolution of the bubble diameter in Figure \ref{fig2}(b).
 The bubbles reached this size within 0.01 s of the laser irradiation, and their size did not appear to change during the laser irradiation. Such vapor-rich bubbles on local heating points are known to oscillate violently at several hundred kilohertz \cite{Namura2020}; however, because the camera exposure time is sufficiently longer than the bubble oscillation period, the bubble appears to be of a constant size. 
When the laser irradiation was stopped, vapor-rich bubbles disappeared more quickly than air-rich bubbles.
Figure \ref{fig2}(c) shows the vapor-rich bubble disappearing captured at 50,000 fps. The time interval between images is 20 \si{\us}.
When the laser irradiation was stopped, the diameter of the vapor-rich bubble decreased from 14 \si{\um} to 5 \si{\um} within 20 \si{\us}.
This rapid shrinkage was mainly due to cooling and condensation of the vapor inside the bubble.
The residual bubble then dissipates over a period of 0.1 second, as shown in Figure \ref{fig2}(f).
This residual bubble is a small amount of air that was contained in the vapor-rich bubble.
The percentage of air molecules in the bubble during laser irradiation is estimated to be less than 10\%.\cite{Hiroshige2024}
 As described in our previous study, the best way to determine whether the inside of a bubble is air-rich or vapor-rich is to measure the bubble's lifetime after the laser irradiation is stopped. \cite{Namura2017} However, under local heating conditions such as those in this experiment, there is a definite difference in bubble size between air and vapor-rich bubbles, as demonstrated in Figures \ref{fig2}(a) and \ref{fig2}(b). Therefore, the distinction between air and vapor-rich bubbles can be determined by observing the diameter of the bubbles several seconds into laser irradiation.

In addition to their size, vapor-rich bubbles exhibit another notable feature. They exhale bubbles with diameters less than 2 \si{\um}. We call this phenomenon ''bubble jet.'' The blue arrows in Figure \ref{fig2}(c) track the bubbles exhaled from the vapor-rich bubble that move away from the vapor-rich bubble at approximately 750 mm s$^{-1}$. 
Such bubble jets are also observed when non-degassed water is locally heated above a certain heat-generation density, as reported by Li et al.\cite{Li2017} The mechanism that produces the bubble jet is almost the same as that in the case of a high heat-generation density.
When vapor-rich bubbles are formed in a non-degassed liquid, they continuously absorb air molecules dissolved in the surrounding fluid. However, vapor-rich bubbles can exhale the air they take in through violent collapse during vibrations and thus are thought to remain vapor-rich.\cite{Li2017}
The fact that the exhaled bubbles were air-rich bubbles was indicated by the fact that they did not disappear immediately after leaving the heating point.

\begin{figure*}[tbp]
\centerline{\includegraphics[bb = 0 0 497 214, width=15cm]{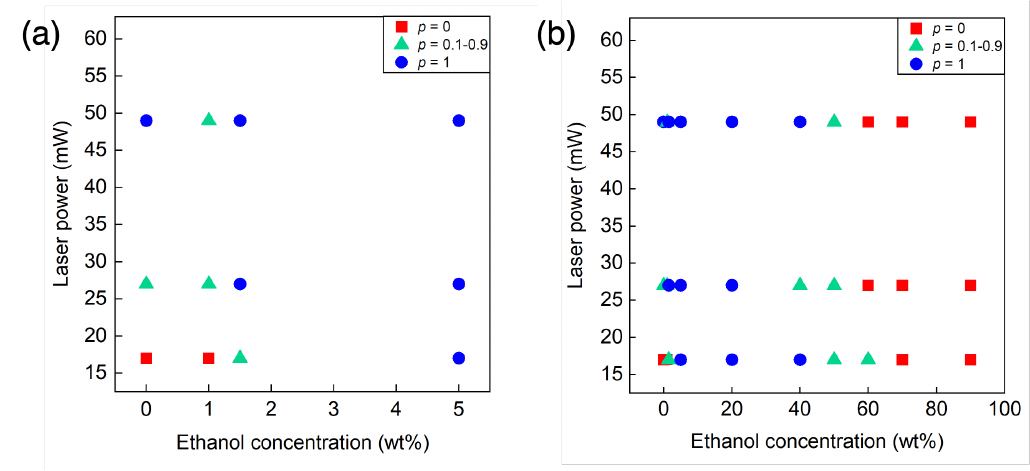}}
\caption{Measured formation probability of vapor-rich bubbles depending on ethanol concentration and laser power. Vertical axis indicates laser power, horizontal axis indicates ethanol concentration, and color and shape of markers indicate generation probability of vapor-rich bubbles. Obtained probability of vapor-rich bubble generation is denoted as $\rho$. Red squares indicate $\rho$ = 0, green triangles indicate $\rho$ = 0.1--0.9, and blue circles indicate $\rho$ = 1. $\rho$ = 1 indicates that vapor-rich bubbles were always generated in ten attempts to generate bubbles. (a) and (b) depict results for ethanol concentrations in 0--5 wt\% and 0--90 wt\%, respectively.}
 \label{fig3}
\end{figure*}

Now, bubbles generated on the local heating point can be roughly classified into air-rich bubbles with a diameter of several tens of micrometers or more and vapor-rich bubbles with a diameter of about 10 \si{\um} and accompanied by bubble jets. Both air and vapor-rich bubbles were observed in ethanol/water mixtures at particular ethanol concentrations. Therefore, we investigated the probability of vapor-rich bubble formation by systematically changing the ethanol concentration and laser power. In Figure \ref{fig3}, the vertical axis indicates the laser power, the horizontal axis indicates the ethanol concentration, and the colors and shapes of the markers indicate the probability of vapor-rich bubble generation.
At each plot point, bubbles were generated ten times under the same conditions, and the number of times a vapor-rich bubble was generated was measured. The laser was irradiated for 3 s, and the bubble size when the laser was stopped was adopted to distinguish between vapor and air-rich bubbles. The probability of vapor-bubble generation is denoted by $\rho$. Red squares indicate $\rho$ = 0, green triangles indicate $\rho$ = 0.1--0.9, and blue circles indicate $\rho$ = 1. $\rho$ = 1 indicates that vapor-rich bubbles were always generated in the ten attempts to generate bubbles. Figure \ref{fig3}(a) presents the results for 0--5 wt\% ethanol/water mixtures. In the case of 0 wt\%, that is, non-degassed pure water, air-rich bubbles were always generated when the laser power was 17 mW, whereas vapor-rich bubbles were always generated when the laser power was 49 mW. In other words, vapor-rich bubbles are generated if the laser power is sufficiently strong, even if water is not degassed. This is consistent with the results of Li et al. \cite{Li2017} The threshold of the laser intensity at which vapor-rich bubbles were generated decreased with increasing ethanol concentration, and above 1.5 wt\%, vapor-rich bubbles were always generated at least once at any laser intensity. 
In the low-laser-power region, giant vapor-rich bubbles were sometimes generated instead of vapor-rich bubbles, as demonstrated in Figure \ref{fig2}(b) (see Supporting Information, Figure S1). The giant vapor-rich bubbles were approximately 100 \si{\um} in diameter, were mainly composed of water vapor, and collapsed approximately 10 \si{\us} after formation.\cite{Wang2018,Detert2020} The intermittent formation of such bubbles is often observed when the laser power density is low.\cite{Namura2020} Because they do not grow into air-rich bubbles, they are classified as vapor-rich bubbles.
Figure \ref{fig3}(b) illustrates the probability of vapor-rich bubble formation for ethanol concentrations in the range of 0--90 wt\%. 
In the range of 1.5--50 wt\%, vapor-rich bubbles were generated at least once at any laser intensity.
In contrast, air-rich bubbles were mainly generated at concentrations greater than 60 wt\%. 
The range of ethanol concentrations that stabilize vapor-rich bubbles is approximately the same as the range of concentrations where solutal-Marangoni forces dominate over thermal-Marangoni forces that we have previously reported (see Supporting Information, Figure S2). \cite{Namura2018} In other words, the vapor-rich bubble is expected to be stabilized by changing the direction of the Marangoni force acting on the exhaled air-rich bubble. Therefore, we decided to observe the movement of air-rich bubbles exhaled from vapor-rich bubbles in more detail.

\begin{figure*}[tbp]
\centerline{\includegraphics[bb = 0 0 712 577, width=15cm]{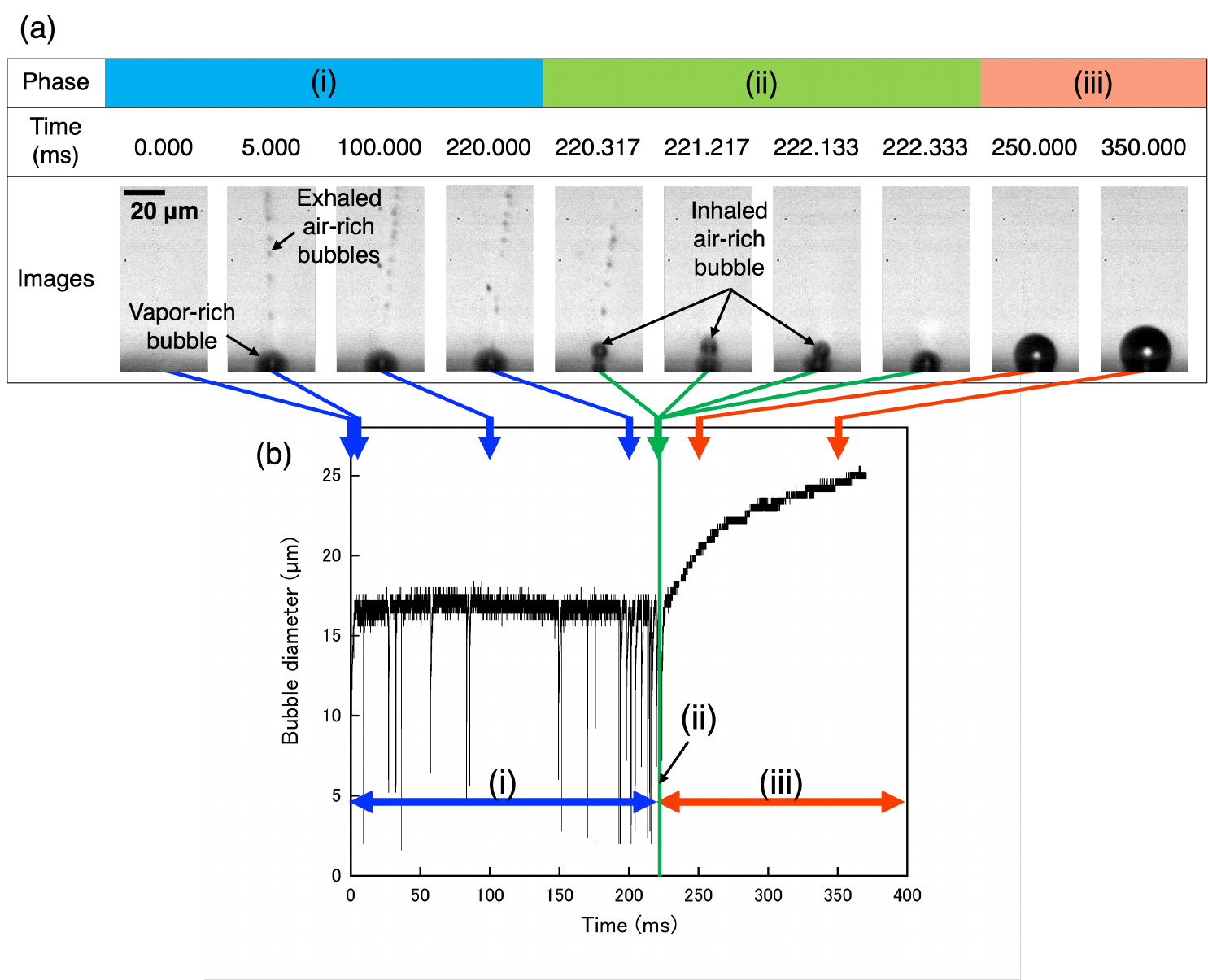}}
\caption{(a) Time series of microscope images during air-rich bubble formation. In Phase (i), vapor-rich bubble is maintained. In Phase (ii), transition from vapor-rich bubbles to air-rich bubbles occurs. In Phase (iii), air-rich bubble grows. (b) Time variation of bubble diameter generated on heating point illustrated in (a). Phases (i)--(iii) correspond to those in (a).}
 \label{fig4}
\end{figure*}

As we have shown in previous studies, in pure water, the round-trip motion of exhaled air-rich bubbles around the vapor-rich bubble initiated the growth of the air-rich bubble on the heat source. \cite{Namura2024}
Therefore, we checked whether the formation of vapor-rich bubbles that exhale air-rich bubbles and the reabsorption of the exhaled air-rich bubbles also occurred in ethanol/water mixtures.
Figure \ref{fig4}(a) illustrates the early stages of air-rich bubble formation at an ethanol concentration of 1 wt\% and a laser power of 17 mW. 
The frame rate was 60000 fps. Laser irradiation was initiated at 0 s. The time evolution of the bubble diameter in Figure \ref{fig4}(a) is depicted by the solid black line in Figure \ref{fig4}(b). Note that this graph illustrates bubble diameters measured from microscope images captured with exposure times longer than the vapor-rich bubble oscillation period and does not reflect bubble oscillations on the order of several hundred kilohertz. The early stages of air-rich bubble formation can be divided into three phases. In Phase (i), immediately after the start of laser irradiation, vapor-rich bubbles of approximately 8 \si{\um} radius with a bubble jet were observed (Figure \ref{fig4}(a), Phase (i)). Figure \ref{fig4}(b) also demonstrates that the size of the bubbles remains constant from 0 to 220 ms, indicating the characteristics of vapor-rich bubbles. Subsequently, in Phase (ii), the exhaled air-rich bubbles clumped together to form a mass with a radius of approximately 5 \si{\um} and stopped near the vapor-rich bubbles (Figure \ref{fig4}(a), Phase (ii)).
This relatively large air-rich bubble is annotated as an ''inhaled air-rich bubble'' in Figure \ref{fig4}(a).
 Subsequently, the air-rich bubble is swallowed by the vapor-rich bubble. This transition occurred within a period of only 2 ms. In Phase (iii), the bubbles began to grow (Figures \ref{fig4}(a) and \ref{fig4}(b), Phase (iii)). 
 Such growth is typical when bubbles incorporate dissolved air in a liquid.\cite{Baffou2014,Namura2017}
The length of Phase (i) tends to increase rapidly at ethanol concentrations above 2 wt\% (see Supporting Information Figure S3), and especially at 5 wt\%, it was observed that the bubbles do not change for more than 11 minutes (see Supporting Information Figure S4). When vapor rich bubbles were stable for long periods of time, the phenomenon of relatively large bubbles returning to the vicinity of the vapor-rich bubbles, as shown in Phase (ii), was not observed.

\begin{figure*}[tbp]
\centerline{\includegraphics[bb = 0 0 614.84 256.8, width=15cm]{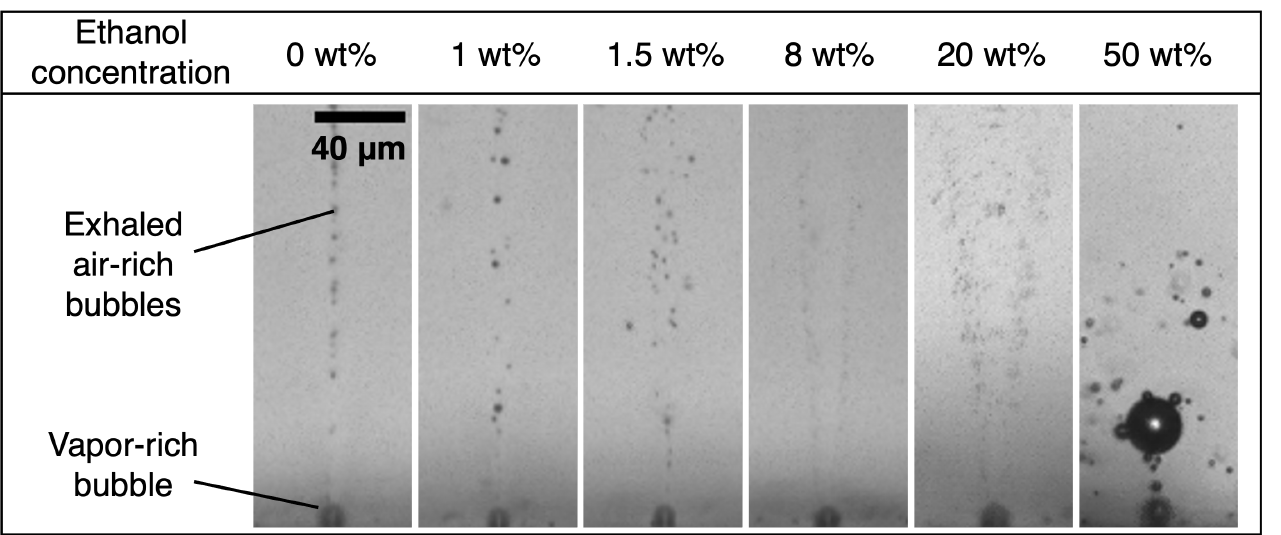}}
\caption{Dependence of the spatial distribution of air-rich bubbles exhaled from vapor-rich bubbles on ethanol concentration.}
 \label{fig5}
\end{figure*}

To better understand the forces acting on the exhaled air-rich bubbles, we examined the spatial distribution of the air-rich bubbles during which the vapor-rich bubble is maintained. Figure \ref{fig5} shows the dependence of the spatial distribution of air-rich bubbles exhaled from vapor-rich bubbles on ethanol concentration. 
Vapor-rich bubbles were generated at the heat source, and tiny air-rich bubbles exhaled from these vapor-rich bubbles moved away from the vapor-rich bubbles. At 0 wt\%, the air-rich bubbles moved away from the heat source along a line perpendicular to the substrate surface. However, the bubbles became unaligned upon addition of ethanol. In particular, at 8--20 wt\%, air-rich bubbles were avoided immediately above the vapor-rich bubbles. At 50 wt\%, the positions of the air-rich bubbles became random. According to our previous study, the solutal-Marangoni forces are significantly strong compared to the thermal-Marangoni forces at 8--20 wt\%, and the bubbles should be driven from the hot to the cold parts. When the vapor-rich bubbles create a strong flow, it is possible that the high-temperature fluid jets out in the region where air-rich bubbles are less abundant at 8--20 wt\% (see Figure S4 of Ref. \cite{Hiroshige2024}). It is most likely that the air-rich bubbles moved to avoid this high-temperature area because of the Marangoni force. However, vapor-rich bubbles cannot guarantee a strong flow in an ethanol/water mixture. If an ethanol concentration gradient is formed on the surface of the vapor-rich bubble, the Marangoni force acting on the vapor-rich bubble will attempt to create a flow from the low-temperature part to the high-temperature part. To understand the movement of exhaled air-rich bubbles, it is necessary to verify the flow created by the vapor-rich bubbles.

\begin{figure*}[tbp]
\centerline{\includegraphics[bb = 0 0 568 302, width=15cm]{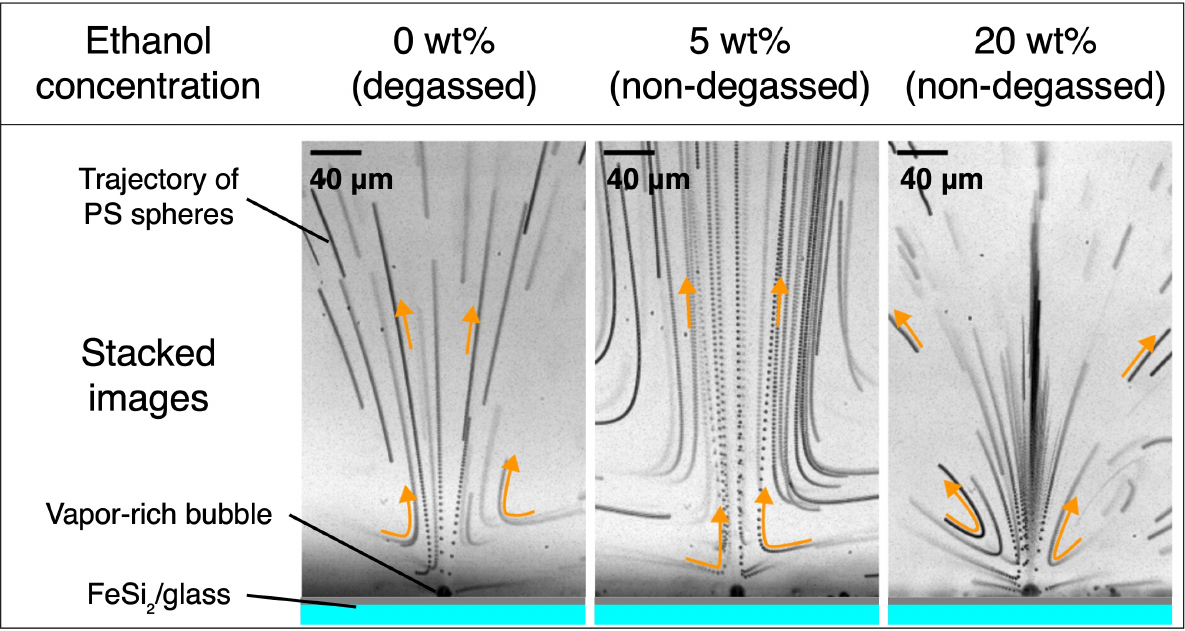}}
\caption{Typical flow observed around vapor-rich bubbles in degassed water and non-degassed ethanol/water mixtures of 5 and 20 wt\%. To visualize flow, 300 images taken in 0.01 s were stacked. Orange arrows indicate direction of flow.}
 \label{fig6}
\end{figure*}

\begin{figure*}[tbp]
\centerline{\includegraphics[bb = 0 0 821 574, width=15cm]{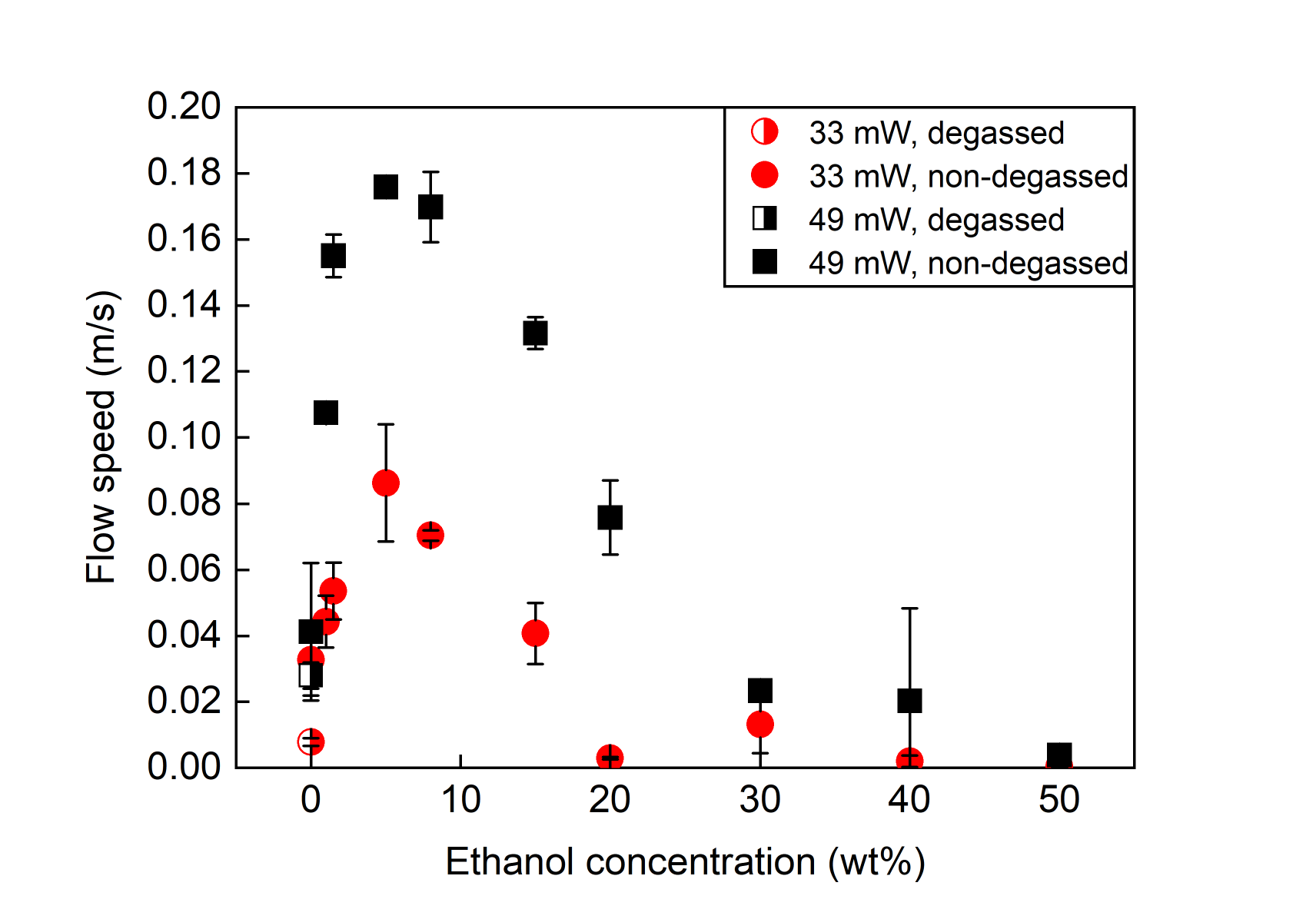}}
\caption{Dependence of flow speed around vapor-rich bubbles on ethanol concentration. Red circles and black squares depict results in non-degassed ethanol/water mixtures at laser powers of 33 mW and 49 mW, respectively. Red/white and black/white markers are results in degassed water. Error bars indicate standard deviation ($n = 3$).}
 \label{fig7}
\end{figure*}

To verify the flow created by the vapor-rich bubble, a solid polystyrene sphere was added to the fluid and its motion was observed. The polystyrene sphere was driven only by the drag force from the flow, because the Marangoni force did not act on the solid-liquid surface. Therefore, the trajectory of the polystyrene sphere is expected to visualize the flow effectively.
Figure \ref{fig6} illustrates the typical vapor-rich bubbles and flow patterns when a 33-mW laser was irradiated on an FeSi$_2$ thin film immersed in degassed water and non-degassed ethanol/water mixtures of 5 and 20 wt\%. 
The images were observed from a direction parallel to the substrate surface and captured approximately 3 s after the start of laser irradiation. The small black circles above the heating point on the substrate represent vapor-rich bubbles. The black dots and lines in the fluid are the PS spheres added to visualize the flow. To make the motion of the PS spheres easier to understand, 300 images taken at 0.01 s were stacked such that the trajectory of the PS spheres appeared as a series of lines and dots. In areas where the PS spheres were moving fast, the spheres captured in each frame appeared to be far apart, whereas, in areas where the PS spheres were moving slowly, the spheres appeared to be strung together like lines.
Orange arrows indicate the direction of movement of the spheres. 
The flow shown in Figure \ref{fig6} is observed when vapor-rich bubbles form on the local heating point. When the photothermally converted thin film was slightly displaced from the beam waist of the irradiating laser to prevent the formation of bubbles, almost no flow was observed (see Supporting Information Figure S5). In other words, the observed flow is not driven directly by light, as in the case of optical streaming \cite{Chraibi2011,Hosokawa2020}, but rather by vapor-rich bubbles.
Contrary to our expectations, the flow speed around the bubbles generated in the 5 wt\% ethanol-water mixture was greater than that around the bubbles generated in degassed water.
Therefore, the relationship between ethanol concentration and flow speed around vapor-rich bubbles was quantitatively investigated. Because the flow speed in the vicinity of vapor-rich bubbles varies significantly depending on the position, the measurement point for the flow speed was set at a distance of 330 \si{\um} from the laser irradiation point on the substrate perpendicular to the substrate surface. The motion of the PS sphere at the measurement point was analyzed employing particle tracking velocimetry to calculate the average speed of the particles passing through the measurement point during a period of 3--3.5 s after the start of laser irradiation. This measurement was repeated thrice under the same conditions. The results are presented in Figure \ref{fig7}. 
The vertical and horizontal axes represent flow speed and ethanol concentration, respectively. The markers in the graph represent the mean values of three measurements, and the error bars are calculated from the standard deviations. The red circles and black squares indicate the experimental results when irradiated with 33 mW and 49 mW laser, respectively. The red/white circles and black/white squares represent the results for degassed water. A comparison of the results at 49 mW reveals that the flow around the vapor-rich bubble was faster in the non-degassed ethanol/water mixture than in degassed water when the ethanol concentration was between 0 and 20 wt\%.
Furthermore, the flow speed reached its maximum when the ethanol concentration was 5 wt\%. At this time, the flow speed in the ethanol/water mixture was 6 times higher than that in the degassed water. 
A similar trend is observed at a laser intensity of 33 mW, where the flow speed in the 5 wt\% ethanol/water mixture was 11 times higher than that of the degassed water.
Moreover, tracing the motion of the PS sphere moving closer to the vapor-rich bubble showed that the flow speed exceeded 1 m s$^{-1}$ even at a distance of 50 \si{\um} from the heating point (see Supporting Information, Figure S6).
Thus, the ethanol/water mixture generates vapor-rich bubbles without degassing and enhances the flow speed created by the vapor-rich bubbles. This is an interesting result from an applicational perspective.
In addition, when isopropanol or 1-butanol is mixed with water instead of ethanol, vapor-rich bubbles can also be generated without degassing the liquid, which has the effect of increasing flow speed (see Supporting Information, Figure S7).

Why does such convection enhancement occur? 
First, we focused on the change in the fluid properties caused by the mixing of ethanol and water. Previous studies reveal that the flow around a vapor-rich bubble is similar to that created by a stokeslet. \cite{Namura2017} Blake et al. described the flow created by a stokeslet in terms of an analytical equation, in which viscosity is the only physical property of the fluid included in the equation. \cite{Blake1974}
The viscosity of the ethanol/water mixture depends on the ethanol concentration; when the ethanol concentration is 40--50 wt\%, the viscosity is approximately twice that of pure water. \cite{Khattab2012}
From the equation given by Blake et al., the flow speed is inversely proportional to the viscosity; therefore, the addition of ethanol should decrease the flow speed. Therefore, the higher flow speed at 5 wt\% could not be explained by a change in the properties of the fluid.
Next, we considered Marangoni flow generated by vapor-rich bubbles. 
Our previous studies have shown that at 5 wt\% ethanol concentration, the solutal-Marangoni forces acting on the air-rich bubble are dominant, generating a flow from the cold to the hot part of the bubble \cite{Namura2018}.
If the same force acts on the vapor-rich bubble, a flow is generated in the direction opposite to that observed. 
This suggests that either the Marangoni force is barely involved in the generation of the vapor-rich bubble flow or that no ethanol concentration gradient occurs at the vapor-rich bubble surface, where intense evaporation and condensation occur. The Marangoni force is not a factor that enhances the flow.
Based on the above results, it is expected that the most important contribution to the generation and enhancement of the flow is the oscillation of the vapor-rich bubble. Even under heating with a CW laser, vapor-rich bubbles oscillate with large amplitudes on the order of several hundred kilohertz \cite{Namura2020}. It is known that when bubbles oscillate, a flow similar to the observed flow is produced \cite{Marmottant2003}. The addition of ethanol may have altered the processes of evaporation and condensation of the liquid, resulting in a change in the amplitude and period of the bubble oscillations. 
To understand the velocity of the flow created by the bubbles, it is necessary to capture changes in the amplitude, frequency, and shape of the bubbles; however, this requires an ultra-high-speed camera of the order of Mfps, which is a future challenge. It is important to note that the flow direction created by the vapor-rich bubbles was approximately the same regardless of the ethanol concentration. Based on these results, the motion of air-rich bubbles exhaled from vapor rich bubbles is discussed qualitatively below.

\begin{figure*}[tbp]
\centerline{\includegraphics[bb = 0 0 561.84 805.2, width=13cm]{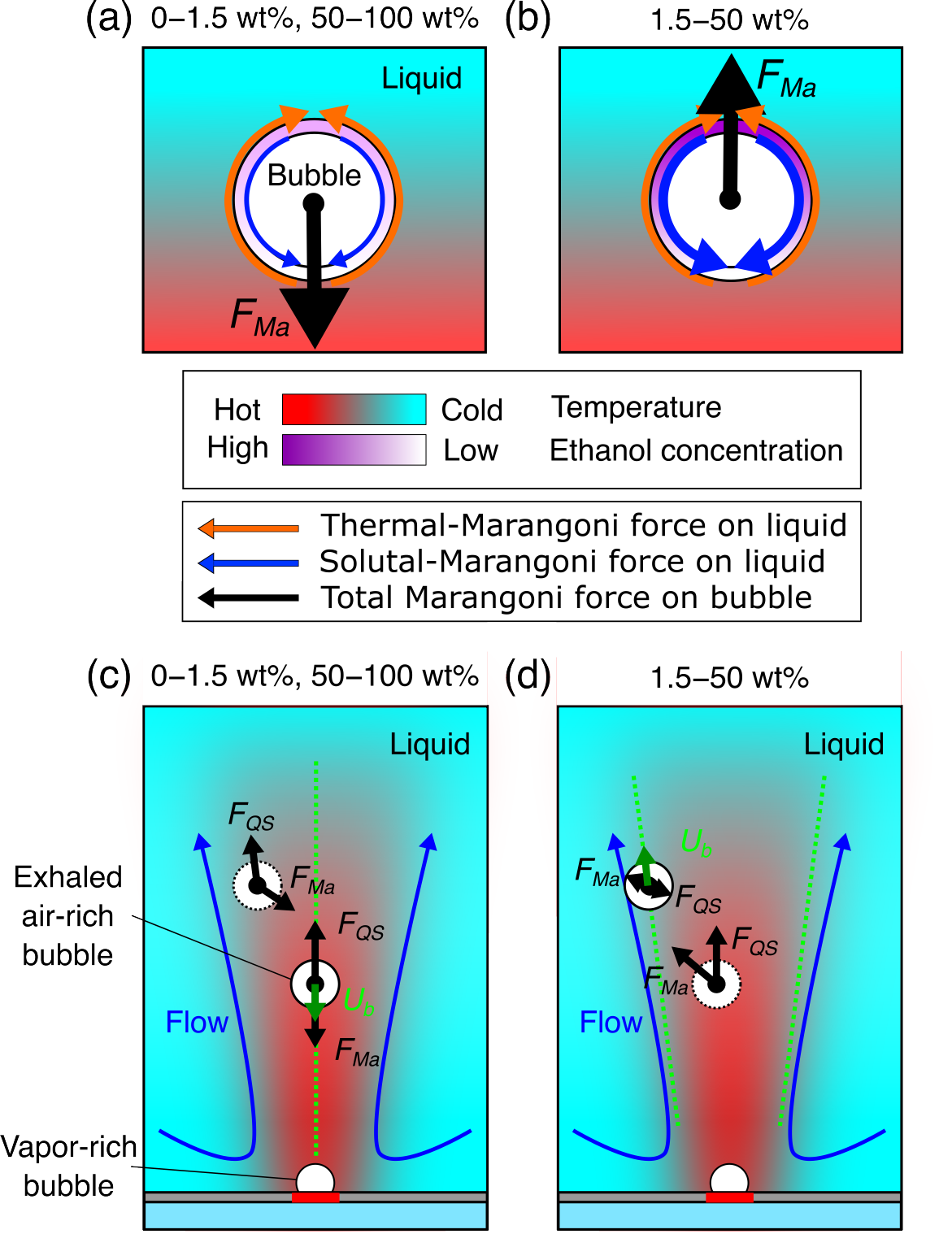}}
\caption{Schematic of direction of thermal- and solutal-Marangoni forces acting on exhaled air-rich bubbles and direction of bubble motion at the ethanol concentration of (a) 0--1.5 wt\% and 50--100 wt\% and (b) 1.5--50 wt\%. Diagram of a qualitative discussion of the balance between the Marangoni force and the quasi-steady drag force acting on the exhaled air-rich bubbles at ethanol concentrations of (c) 0--1.5 wt\% and 50--100 wt\% and (d) 1.5--50 wt\%.}
 \label{fig8}
\end{figure*}

According to our previous studies, the forces governing the motion of an exhaled air-rich bubble are primarily the Marangoni force and quasi-steady drag force.\cite{Namura2024} We now consider the direction of these two forces. First, we considered the Marangoni force acting on an air-rich bubble. As shown in Figures \ref{fig8}(a) and \ref{fig8}(b), the air-rich bubble exposed to a temperature gradient in an ethanol-water mixture is subjected to thermal-Marangoni and solutal-Marangoni forces. The red and blue gradients represent temperature gradients. The surface tension of the bubble weakened at high temperatures, and as indicated by the orange arrows, a thermal-Marangoni force was applied to the surrounding fluid from the high-temperature region to the low-temperature region. The temperature gradient also induced an ethanol concentration gradient on the bubble surface, as indicated by the purple and white gradients. This was because ethanol is more volatile than water. Because water has stronger surface tension than that of ethanol, solutal-Marangoni forces are induced from a low-temperature area with more ethanol to a high-temperature area with less ethanol. According to our previous studies, the thermal-Marangoni forces are stronger than the solutal-Marangoni forces in the 0--1.5 and 50--100 wt\% ethanol concentration ranges (see Supporting Information, Figure S2). \cite{Namura2018} As a result, the reaction of the total Marangoni force drives the bubbles from the low temperature region to the high temperature region, as indicated by the black arrows (Figure \ref{fig8}(a)). By contrast, for ethanol concentrations in the range of 1.5--50 wt\%, the ethanol concentration gradient on the bubble surface was larger, resulting in stronger solutal-Marangoni forces than the thermal-Marangoni forces. Consequently, the bubbles were driven from the hotter to the colder region (Figure \ref{fig8}(b)). We then consider bubble motion by considering the balance between the total Marangoni force acting on the bubble and quasi-steady drag force.

First, we consider the motion of air-rich bubbles at 0--1.5 wt\% and 50--100 wt\% (Figure \ref{fig8}(c)). The vapor-rich bubbles generated at the heat source created the flow indicated by the blue arrows. Therefore, the hot fluid spread in the direction perpendicular to the substrate surface, as indicated by the red color in Figure \ref{fig8}(c). \cite{Hiroshige2024} We assume that an air-rich bubble exists at the position indicated by the white circle surrounded by the solid black line and moves with velocity $\bm{U_b}$. The flow velocity of the surrounding liquid was $\bm{U}$. The quasi-steady drag force $\bm{F_{QS}}$ acting on the bubble is in the direction $\bm{U}-\bm{U_b}$ and its strength is approximately proportional to the magnitude of $\bm{U}-\bm{U_b}$. If the bubble is above the green dashed line, $\bm{F_{QS}}$ acts away from the heat source. However, the total Marangoni force $\bm{F_{Ma}}$ acts toward the heat source. Considering that the bubble mass is almost negligible, the bubble velocity $\bm{U_b}$ changes such that $\bm{F_{Ma}}$ and $\bm{F_{QS}}$ are balanced. That is, if $\bm{F_{Ma}}$ is weak, $\bm{F_{QS}} \sim 0$, or $\bm{U}-\bm{U_b} \sim 0$. If $\bm{F_{Ma}}$ is strong, $\bm{U}-\bm{U_b}$ is large. To satisfy this condition, $\bm{U_b}$ is in the direction of the heat source, and the bubble moves back toward the heat source. This phenomenon was also observed in our previous study. \cite{Namura2024}
In our previous studies, $\bm{F_{Ma}}$ in pure water was found to be sufficiently strong to pull air-rich bubbles back toward the heat source against the flow when the exhaled air-rich bubbles fuse with each other and exceed a radius of a few micrometers.
As ethanol was added to water, the total Marangoni force toward the heat source decreased. In other words, the air-rich bubbles are less likely to return to the heat source. Therefore, vapor-rich bubbles are more likely to form as the ethanol concentration increases in the 0--1.5\% ethanol range. This is also consistent with the fact that the formation of air-rich bubbles is enhanced in the 50--100 wt\% ethanol concentration range, where the Marangoni force toward the heat source is stronger.
What if an air-rich bubble was outside the green line? Assuming that the air-rich bubble was stationary at the position indicated by the white circle surrounded by the black dashed line (Figure \ref{fig8}(c)), this bubble will be subjected to a quasi-steady drag force in the direction of the flow speed $\bm{U}$. The bubble was also subjected to $\bm{F_{Ma}}$ in the direction from low to high temperatures. 
Consequently, the bubble was attracted to the direction of the dashed green line and moved along this line. The stronger the Marangoni force, the stronger the constraining force on the position of the green dashed line. At 0--1.5 wt\%, the bubbles do not align in a straight line as more ethanol is added because the total Marangoni force becomes weaker (Figure \ref{fig5}).

Next, we consider the range of ethanol concentrations from 1.5--50 wt\% (Figure \ref{fig8}(d)). First, we assume that the air-rich bubble is stationary at the position indicated by the white circle surrounded by the black dashed line. Considering the direction of the flow created by the vapor-rich bubble, $\bm{F_{QS}}$ forced the bubble away from the heat source. In addition, $\bm{F_{Ma}}$ acts in the direction from the high-temperature region to the low-temperature region. Therefore, the air-rich bubble was subjected to a force in the direction away from the heat source and from the flow of the hot fluid directly above the vapor-rich bubble. Consequently, the air-rich bubble was driven to a position where the temperature gradient was relatively low, as indicated by the dashed green lines. Consider an air-rich bubble in motion at the position indicated by the white circle surrounded by the solid black line. This bubble was subjected to a weak Marangoni force in the direction from the high-temperature to the low-temperature region. $\bm{F_{QS}}$ balances the Marangoni forces. However, as $\bm{F_{Ma}}$ is small, $\bm{F_{QS}} \sim 0$, implying that $\bm{U} \sim \bm{U_b}$. Finally, the air-rich bubble moves along the flow created by the vapor-rich bubble at the position indicated by the green dashed line. This is consistent with the observations at 8 and 20 wt\% in Figure \ref{fig5}. In this range, no force acts on the air-rich bubble in the direction back to the heat source unless the exhaled air-rich bubble enters the flow in the direction of the vapor-rich bubble. In other words, for ethanol concentrations in the 1.5--50 wt\% range, exhaled air-rich bubbles were rarely reabsorbed by vapor-rich bubbles. Therefore, it can be said that the addition of ethanol manipulates the direction of the Marangoni force acting on the exhaled air-rich bubbles and succeeds in stabilizing the vapor-rich bubbles. The method reported in this paper can stabilize vapor-rich bubbles at low heat generation densities without degassing the fluid. Furthermore, the method successfully enhanced the flow around the vapor-rich bubbles. These results are expected to be useful for pumping fluids in microchannels.

\section{CONCLUSION}
We investigated the formation of vapor-rich bubbles and the surrounding flow when an ethanol/water mixture was heated locally. The results reveal that vapor-rich bubbles were stably generated when the ethanol/water mixture's concentration was 1.5--50 wt\%, even though the liquid was not degassed. 
In this concentration range, vapor-rich bubbles were formed even at relatively low heat generation densities that result in the growth of an air-rich bubble on the heat source in pure-water. This indicates that vapor-rich bubbles can be stabilized even when the fluid is gradually heated.
Vapor-rich bubbles generated in the non-degassed ethanol/water mixtures exhaled air-rich bubbles. The key to stabilizing vapor-rich bubbles is determining whether these exhaled air-rich bubbles return to vapor-rich bubbles above the heating point. The motion of the exhaled air-rich bubbles can be qualitatively explained by the balance between the thermal and solutal-Marangoni forces and the quasi-steady drag force acting on them.
In the range of ethanol concentrations from 1.5 to 50 wt\%, the combined thermal and solutal-Marangoni forces pulled the exhaled air-rich bubbles from the hot to the cold regions. Furthermore, it was experimentally confirmed that in this ethanol concentration range, the Marangoni force did not significantly affect the vapor rich bubbles, causing a flow from hot to cold regions on the vapor rich bubbles. This flow moves the exhaled air-rich bubbles away from the vapor-rich bubbles. These effects caused the air-rich bubbles to lose the opportunity to fuse with vapor-rich bubbles, resulting in the stabilization of the vapor-rich bubbles. Furthermore, measurements of the flow created by the vapor-rich bubbles showed that for ethanol concentrations between 0 and 20 wt\%, the flow around the bubbles was faster in the non-degassed ethanol/water mixture than that in degassed water. The flow speed reached its maximum when the ethanol concentration was 5 wt\%, and the flow speed was 6--11 times higher than that of degassed water.
Utilizing ethanol/water mixtures allows vapor-rich bubbles to be generated without degassing and increases the flow speed produced by the vapor-rich bubbles. Vapor-rich bubbles generated in ethanol/water mixtures have potential applications in driving and mixing fluids in micro heat exchangers.

\section{EXPERIMENTAL SECTION}
\subsection{Preparation of FeSi$_2$ thin films}
We prepared an amorphous FeSi$_2$ thin film for photothermal conversion. The film was prepared on a glass substrate adopting radio-frequency magnetron sputtering. An FeSi$_2$ target (purity:99.9\%) was adopted as the sputtering source material, and the distance between the target and the substrate was $\sim$90 mm. The deposition chamber was evacuated to a vacuum of $> 2.5 \times$ 10$^{-5}$ Pa. Subsequently, Ar (99.99\%) was introduced as the sputtering gas at a rate of 9 sccm. The sputtering was performed at room temperature. During sputtering, the total gas pressure was maintained at 0.86 Pa, and the substrate was rotated rapidly. Subsequently, an FeSi$_2$ layer of 50 nm was deposited on a glass substrate at a deposition rate of 0.06 nm s$^{-1}$. The optical reflectance and transmittance of the prepared thin films were measured with a single-beam spectrophotometer and an integration sphere (ISP-REF; Ocean Optics). The optical transmittance and absorption of the film were determined to be 0.13 and 0.33, respectively, at a wavelength of 785 nm, where the photothermal conversion experiments were conducted (see Supporting Information, Figure S8).

\subsection{Preparation of fluidic cell}
The prepared FeSi$_2$ thin film was placed in a UV ozone cleaner (UV253S, Filgen) for 60 min to improve surface wettability. The cleaned thin films were placed in a glass cell (10 mm $\times$ 10 mm $\times$ 58 mm, F15-G-10, GL Science) and fixed with a magnet. The glass cell was then filled with non-degassed water, degassed water, or a non-degassed ethanol/water mixture. 
The fluid cell was filled with fluid carefully to prevent air-rich bubbles from entering the cell and closed with a lid.
For the non-degassed water, ultrapure water (18.2 M\si{\ohm} cm from Millipore-Direct Q UV3, Merck) was utilized. To prepare degassed water, ultrapure water was sonicated under vacuum ($\sim$3 kPa at 25 \si{\degreeCelsius}) for 20 min. An ethanol/water mixture was prepared by mixing non-degassed ultrapure water and ethanol (FUJIFILM Wako Pure Chemical Corporation). When conducting the experiments to measure the flow speed around the bubble, PS spheres with diameters of 1.9 \si{\um} (R0200, Thermo Scientific) were dispersed in the fluid to visualize the flow. The particle number density of the polystyrene spheres was set to $\sim$2 $\times$ 10$^6$ cm$^{-3}$.

\subsection{Observation of microbubbles and flow generation}
The prepared fluidic cell was attached to the experimental setup illustrated in Figure \ref{fig1}. 
A continuous-wave laser with a wavelength of 785 nm was focused on the FeSi$_2$ thin film by employing objective Lens 1 (20$\times$, NA=0.40). The shape and size of the laser spot on the thin film were adjusted with objective Lens 2 (10$\times$, NA = 0.26) and Camera 1 (HXC20, Baumer), respectively. The diameter of the laser spot on the FeSi$_2$ thin film was approximately 3.8 $\pm$ 0.2 \si{\um}. 
The laser power at the sample surface was tuned from 0 to 50 mW. 
Most of the light absorbed by the FeSi$_2$ thin film is converted into heat, and the laser spot on the film acts as a heater that generates bubbles in the liquid.
The generated bubbles were observed with Camera 2 (FASTCAM Mini AX50, Photron) with objective Lens 3 (10$\times$, NA=0.26/ 20$\times$, NA=0.40/ 50$\times$, NA=0.42). The frame rate of the camera was set to 30000 or 60000 fps. A short-pass filter was placed before the camera to remove the 785-nm laser light.

\subsection{Measurement of flow speed}
A particles tracking velocimetry (PTV) software (Flow Expert2D2C, Kato Koken) was used to measure the flow speed. As described in Section B, the PS spheres were dispersed in the fluid and used as tracer particles. The images used to measure the flow speed were captured by camera 2 in Figure \ref{fig1}, using a series of approximately 15,000 images captured at 30,000 fps. Images were obtained 3-s after the start of laser irradiation. The observation plane was perpendicular to the direction of gravity and was not significantly affected by gravity. The depth of field of the microscope was approximately 50 \si{\um}, and the particles clearly observed in the images were located within this range. The measurement point for the flow speed was set 330-\si{\um} from the laser irradiation point on the substrate, perpendicular to the substrate surface. The flow speed was obtained by averaging the velocities of all the particles passing through the 15-\si{\um} square area centered on the observation point. The flow speed was measured three times at the same laser power and ethanol concentration, and the mean and standard deviations were calculated.

\begin{acknowledgments}
This study was supported by JSPS KAKENHI Grant Nos. 19K21932 and 21H01784 and the JST FOREST Program (Grant Number JPMJFR203N, Japan). 
It was also financially supported by a collaborative research project between Kyoto University and Mitsubishi Electric Corporation on Evolutionary Mechanical System Technology. 
\end{acknowledgments}

\bibliography{2017butanol}

\end{document}